# A Local View of the Observable Universe

MARCO BRUNI,[1,2,3] SABINO MATARRESE,[4] and ORNELLA PANTANO[4]

[1] *Astronomy Unit, School of Mathematical Sciences,*
*Queen Mary & Westfield College, Mile End Road, E1 4NS London, UK*

[2] *Dipartimento di Astronomia, Università di Trieste, via Tiepolo 11, 34131 Trieste, Italy*

[3] *SISSA, via Beirut 2-4, 34013 Trieste, Italy*

[4] *Dipartimento di Fisica "Galileo Galilei",*
*Università di Padova, via Marzolo 8, 35131 Padova, Italy*

(July 15, 1994)

## Abstract

We present results on the non-linear dynamics of inhomogeneous cosmological models with irrotational dust and a positive cosmological constant, considering, in particular, a wide class with vanishing magnetic Weyl tensor. For those patches of the universe that do not recollapse to a singularity we find a unique attractor, representing a de Sitter vacuum phase. For the (re-)collapsing regions we find a family of (Kasner) attractors, so that generically these regions fall in spindle-like singularities. These results give substantial support to the idea that the universe can be very inhomogeneous on ultra-large, super-horizon scales, whit observers living in those (almost) isotropic regions that emerge from an inflationary phase.

Typeset using REVTEX

Cosmology in the XX century has been essentially based on the 2 (+1) parameter ($H_0$, $\Omega_0$, and $q_0$) standard Friedman-Robertson-Walker (FRW) models. In 1981 inflation came on the scene as a possible solution to the conundrums of the big-bang FRW scenario [1,2]. However, despite the fact that these flaws are in a way or another related to the question FRW models cannot answer: *why the observable universe looks isotropic*, and the inflationary scenario was proposed also to answer this question, in practice most of the work on inflation has been done in the framework of isotropic FRW models. It was indeed soon recognised that during an inflationary phase initially present *small perturbations* are swept away, a fact that led to the conjecture that this could be the signature for a more general property of inflation, going under the name of cosmic no-hair theorem [3]. Roughly speaking, inflation should erase previously present inhomogeneities, leaving us with a unique possible observable universe: the isotropic one. Investigations to prove some restricted version of this general conjecture were done first in the framework of homogeneous anisotropic models, [4] then also considering inhomogeneous spacetimes, [5] but with practical examples limited to geometries with some degree of symmetry (e.g. [6,2] and references therein). The dominant perspective emerging from these analysis is that in a universe model either inflation occurs everywhere, or there is no inflation at all. Instead, with truly inhomogeneous initial conditions, one can expect that there will always be patches of the universe that will inflate and isotropize, while others will not, eventually recollapsing to a singularity. In this perspective, we propose that a weak cosmic no-hair theorem holds.

It is the aim of this Letter [7] to present some results about the evolution of inhomogeneous universes with irrotational dust of density $\rho$ and a positive cosmological constant $\Lambda$, in particular, studying the dynamics of the wide class of models with vanishing magnetic part of the Weyl tensor, $H_{ab} = 0$.

These spacetimes were first considered in [8], while their first cosmological implementation was given in [9], where it was also shown that for the case of dust their time evolution is given by a system of six first-order ordinary differential equation for $\rho$ (matter density), $\Theta$ (expansion scalar), $\sigma_1$ and $\sigma_2$ (two independent eigenvalues of the traceless shear tensor $\sigma_{ab}$), and $E_1$ and $E_2$ (eigenvalues of the traceless electric Weyl tensor $E_{ab}$). Thus, the evolution of each "fluid element" in the $H_{ab} = 0$ models is purely local once initial condition satisfying the appropriate constraint equations [8] are given. Because of this property we dubbed these models silent universes [10,11]. As shown in [10], $H_{ab}$ is a good approximation (at least at second order in perturbations of a FRW background) outside the Hubble horizon, where also pressure gradients can be neglected, so that these models may provide a fair picture of the universe on ultra-large scales. In other words, the time evolution of each super-horizon sized "volume element" of the universe is well described by the equations of the $H_{ab} = 0$ models.

The dynamics of these models with $\Lambda = 0$ has been presented in detail elsewhere [11]. Here we only point out that they put the flatness problem [1,2,12] in a rather unusual perspective: indeed, independently of its initial value, $\Omega$ ultimately tends to zero, both for expansion and collapse, with most of the universe volume dominated by expanding voids. Defining $\sigma_\pm = \frac{1}{2}(\sigma_1 \pm \sigma_2)$ and $E_\pm = \frac{1}{2}(E_1 \pm E_2)$ the dynamics of irrotational dust with $H_{ab} = 0$ and $\Lambda > 0$ is given by

$$\dot{\rho} = -\Theta\,\rho \;, \tag{1a}$$

$$\dot{\Theta} = -\frac{1}{3}\Theta^2 - 2\,\sigma^2 - \frac{1}{2}\rho + \Lambda \;, \tag{1b}$$

$$\dot{\sigma}_+ = \sigma_+{}^2 - \frac{1}{3}\sigma_-{}^2 - \frac{2}{3}\Theta\,\sigma_+ - E_+ \;, \tag{1c}$$

$$\dot{\sigma}_- = -2\,\sigma_+\,\sigma_- - \frac{2}{3}\Theta\,\sigma_- - E_- \;, \tag{1d}$$

$$\dot{E}_+ = \sigma_-\,E_- - 3\,E_+\,\sigma_+ - \Theta\,E_+ - \frac{1}{2}\rho\,\sigma_+ \;, \tag{1e}$$

$$\dot{E}_- = 3\,\sigma_-\,E_+ + 3\,E_-\,\sigma_+ - \Theta\,E_- - \frac{1}{2}\rho\,\sigma_- \;, \tag{1f}$$

where $\sigma^2 = 3\,\sigma_+{}^2 + \sigma_-{}^2$ is the shear magnitude. Setting $\sigma_- = E_- = 0$ one obtains the dynamics of a set of universe models generalizing those of Szekeres to the case of $\Lambda > 0$; some of these models were explicitly given in [13]. [14] Equations (1a) and (1b) are actually more general than for the case $H_{ab} = 0$, as they hold for irrotational dust in general. Also, in this case the curvature of the 3-surfaces orthogonal to the matter 4-velocity [15] $u^a$ is

$$^{(3)}R = -\frac{2}{3}\Theta^2 + 2\,\sigma^2 + 2\,\rho + 2\,\Lambda \;. \tag{2}$$

Proofs of cosmic no-hair theorems for homogeneous [4] and inhomogeneous [5] spacetimes were essentially based on the assumption $^{(3)}R \leq 0$, although inflation has been found also in some $^{(3)}R > 0$ models [6,2,16]. For $^{(3)}R < 0$, it follows that $\Theta \geq \sqrt{3\Lambda}$, while $\dot{\Theta} \leq \Lambda - \frac{1}{3}\Theta^2 \leq 0$, so that if at $t_*$ (an arbitrary initial time) $\Theta_* > 0$ (the model is initially expanding), then the model expands forever, with $\Theta$ squeezed between the lower and upper bound and $\Theta \to \sqrt{3\Lambda}$ exponentially, thus approaching (at least locally) a de Sitter spacetime in a timescale $\alpha = \sqrt{3/\Lambda}$.

Let us now drop the assumption $^{(3)}R \leq 0$ and make some general (i.e. not restricted to the $H_{ab} = 0$ case) remarks. From (1b) and $\Theta_* > \sqrt{3\Lambda}$ the upper bound

$$\frac{\Theta(t,\vec{x})}{\sqrt{3\Lambda}} \leq \coth\left[\sqrt{\frac{\Lambda}{3}}(t - t_*) + \coth^{-1}\left(\frac{\Theta_*(\vec{x})}{\sqrt{3\Lambda}}\right)\right] \tag{3}$$

still holds, as in [4,5], but for $\Theta_* < \sqrt{3\Lambda}$ this becomes

$$\frac{\Theta(t,\vec{x})}{\sqrt{3\Lambda}} \leq \tanh\left[\sqrt{\frac{\Lambda}{3}}(t - t_*) + \tanh^{-1}\left(\frac{\Theta_*(\vec{x})}{\sqrt{3\Lambda}}\right)\right] \;. \tag{4}$$

Defining as usual [17,11] by $3\dot{\ell}/\ell = \Theta$ the local scale factor $\ell$, from (1a) one has that $\rho \to 0$ as long as the expansion proceeds, and in this case from (1b) and (3), (4) one gets that $\Theta \to \sqrt{3\Lambda}$ if also anisotropy is suppressed, i.e. $\sigma \to 0$. However, $\Theta$ is no more bounded from below in the general case, so that in general one can also expect recollapse, *even after an inflationary phase*. From (1b) and (2) we see that if $\Lambda \geq \frac{1}{3}\Theta^2$ then $^{(3)}R > 0$, and also if $\dot{\Theta} > 0$ then $^{(3)}R > 0$, while if $\Lambda \leq \frac{1}{3}\Theta^2$ then $\dot{\Theta} < 0$. In particular, this implies that $\Theta < -\sqrt{3\Lambda}$ is a no-return region: every trajectory entering it will undergo collapse with $^{(3)}R > 0$ sufficiently close to the singularity. But if a patch of the universe starting with $\Theta < \sqrt{3\Lambda}$ has to inflate and isotropize, then in order to get the $\Theta \to \sqrt{3\Lambda}$ asymptotic value it has to *super-inflate* [18], with $\dot{\Theta} > 0$ *and* $^{(3)}R > 0$.

In terms of the usual density parameter $\Omega = \Omega_M + \Omega_V$, with the matter and vacuum density parameters defined as usual, $\Omega_M = 3\rho/\Theta^2$ and $\Omega_V = 3\Lambda/\Theta^2$, it follows from (2) that $\Omega > 1 \Rightarrow {}^{(3)}R > 0$ and $^{(3)}R < 0 \Rightarrow \Omega < 1$, and from (1b) that $\dot{\Theta} > 0 \Rightarrow \Omega > 1$, but the reverse relations do not hold, because of the shear.

Now, let us consider the specific case of the dynamics of the $H_{ab} = 0$ models, described by system (1). For $\Lambda = 0$ this system has a single stationary point, [19] given by the origin in phase-space, $\rho = \Theta = \sigma_\pm = \varepsilon_\pm = 0$, and corresponding to Minkowski spacetime. For $\Lambda > 0$ this point bifurcates in 10 new points. Of these, 3 are static ($\Theta = 0$), with only one physically meaningful (the other 2 have $\rho < 0$), and represents the Einstein universe, with $\rho = 2\Lambda$ and $\sigma_\pm = E_\pm = 0$. Another point represents the spatially flat de Sitter spacetime, with $\Theta = \sqrt{3\Lambda}$ (assuming expansion), and $\rho = \sigma_\pm = E_\pm = 0$. Two other points have $\sigma_- = E_- = 0$, thus are degenerate and represent an oblate and a prolate configuration expanding exponentially with $\Theta = \sqrt{\Lambda/3}$ and $\Theta = \sqrt{\Lambda}$ respectively. Finally, the two other pairs of stationary points are physically equivalent replicas of these two latter. Now, the linearized stability analysis of these stationary points shows that all of them are saddle points, thus unstable, except the one representing de Sitter, which is asymptotically stable if $\Theta > 0$, i.e. de Sitter *is the unique attractor for expansion*. It then follows that those patches of the $H_{ab} = 0$ models that do not recollapse expand toward a de Sitter phase and isotropize, i.e. $\sigma_\pm \to 0$ and $E_\pm \to 0$, with $\Omega \to 1$ and $^{(3)}R \to 0$. Therefore, even in case the portion of the universe undergoing inflation was initially negligibly small, in a timescale of order $\alpha = \sqrt{3/\Lambda}$ most of the universe volume will be isotropized by a phase of de Sitter-like inflation: a picture which is in the spirit of chaotic inflation [1]. According to the inflationary scenario, this de Sitter period will be ended by a suitable reheating process, after which a standard FRW phase will occur with $\Omega$ close to unity.

Another view of the phase-space for the $H_{ab}$ models is achieved using $\Omega_M$ and $\Omega_V$, together with the other dimensionless variables [20] $\Sigma_\pm = \sigma_\pm/\Theta$ and $\varepsilon_\pm = E_\pm/\Theta^2$. Obviously these variables diverge for $\Theta = 0$, so that one has to separately consider the $\Theta > 0$ and $\Theta < 0$ cases. Also, it is convenient to introduce a new "time" $\tau = \pm 3 \ln \ell$, using the minus (plus) for the $\Theta < 0$ ($\Theta > 0$) case, so that $d\tau/dt > 0$ in both cases.

Denoting by a prime the derivative with respect to $\tau$, by $\Sigma^2 = 3\,\Sigma_+{}^2 + \Sigma_-{}^2$ the magnitude of the dimensionless shear, and by $\Omega_G = \Omega_M - 2\Omega_V$ the effective gravitational mass density parameter, the evolution equations for our variables for $\Theta < 0$ read

$$\Theta' = \frac{\Theta}{6}(2 + 12\,\Sigma^2 + \Omega_G)\,, \tag{5}$$

$$\Omega_V' = -\frac{\Omega_V}{3}(2 + 12\,\Sigma^2 + \Omega_G)\,, \tag{6a}$$

$$\Omega_M' = -\frac{\Omega_M}{3}(12\,\Sigma^2 - 1 + \Omega_G)\,, \tag{6b}$$

$$\Sigma'_+ = \frac{\Sigma_+}{6}(2 - 12\,\Sigma^2 - 6\,\Sigma_+ - \Omega_G) + \frac{1}{3}\Sigma_-{}^2 + \varepsilon_+ \;, \tag{6c}$$

$$\Sigma'_- = \frac{\Sigma_-}{6}(2 - 12\,\Sigma^2 + 12\,\Sigma_+ - \Omega_G) + \varepsilon_- \;, \tag{6d}$$

$$\varepsilon'_+ = \frac{\varepsilon_+}{3}(1 - 12\,\Sigma^2 + 9\,\Sigma_+ - \Omega_G) - \Sigma_- \,\varepsilon_- + \frac{1}{6}\Sigma_+ \,\Omega_M \;, \tag{6e}$$

$$\varepsilon'_- = \frac{\varepsilon_-}{3}(1 - 12\,\Sigma^2 - 9\,\Sigma_+ - \Omega_G) - 3\,\Sigma_- \,\varepsilon_+ + \frac{1}{6}\Sigma_- \,\Omega_M \;. \tag{6f}$$

We see that now the Raychaudhuri equation (5) is decoupled from the rest of the system (6), i.e. all other equations do not depend on $\Theta$. The advantage we get in introducing system (6) is that, in addition to the non-static ($\Theta \neq 0$) stationary points of system (1), we also have a set of stationary points representing models for which the variables in (1) diverge , while those in (6) obviously have finite constant values. More precisely, system (6) admits many isolated stationary points, 3 of which are unphysical ($\Omega_M = -3$), with many degenerate physically equivalent triplets, and 2 physically equivalent sets of points parametrized by $\Sigma_+$. The physically significant and distinct points are listed in Table I; points L I, L II and L III have $\Omega_V \neq 0$ and are the same non-static points of system (1) described above. Point D I-D VI are degenerate points with $\Omega_V = 0$ (i.e. for finite $\Theta$ they represent $\Lambda = 0$ models), and the set T III represents triaxial configurations parametrized by $\Sigma_+$ (see the notes in Table I).

The linearized stability analysis shows again that for expansion $\Theta > 0$ point L I, locally representing a flat de Sitter universe, is again the unique attractor, while for collapse, $\Theta < 0$, the set T III, together with its conjugate $\overline{\text{T III}}$ (given by the sign exchange $\Sigma_- \to -\Sigma_-$ and $\varepsilon_- \to -\varepsilon_-$ in the expressions in Table I), is attracting. If collapse occurs $\Theta \to -\infty$ and $\Omega_V \to 0$, while $\Sigma_\pm$ and $\varepsilon_\pm$ tend to finite values, and $\Omega_M \to 0$ as well: it can be shown [11] that the sets T III and $\overline{\text{T III}}$ are locally equivalent to the Kasner models, thus the outcome of the stability analysis for system (6) is that collapsing configurations generically tend to a triaxial Kasner-like spindle singularity, with matter and $\Lambda$ having no effects in the final stage.

TABLE I. Stationary points of system (6) (only the physically interesting points are listed). For expansion ($\Theta > 0$) only the de Sitter point L I is asymptotically stable, and for collapse ($\Theta < 0$) the only stable points are those of the Kasner set, i.e. the T III family and its conjugate $\overline{\text{T III}}$. All other points are saddles. Models marked with M are equivalent to Minkowski.

| Point | $\Omega_M$ | $\Omega_V$ | $\Sigma_+$ | $\Sigma_-$ | $\varepsilon_+$ | $\varepsilon_-$ | Model |
|---|---|---|---|---|---|---|---|
| L I | 0 | 1 | 0 | 0 | 0 | 0 | *de Sitter* |
| L II | 0 | 3 | $-1/3$ | 0 | $1/3$ | 0 | *prolate* |
| L III | 0 | 9 | $2/3$ | 0 | 0 | 0 | *oblate* |
| D I | 1 | 0 | 0 | 0 | 0 | 0 | *Flat FRW* |
| D II | 0 | 0 | 0 | 0 | 0 | 0 | *Milne* (M) |
| D III | 0 | 0 | $1/6$ | 0 | 0 | 0 | *Szekeres* (M) |
| D IV | 0 | 0 | $-1/3$ | 0 | 0 | 0 | *Kasner* (M) |
| D V | 0 | 0 | $1/3$ | 0 | $2/9$ | 0 | *Kasner* |
| D VI | 0 | 0 | $-1/12$ | 0 | $1/32$ | 0 | *Szekeres* |
| T III | 0 | 0 | $\Sigma_+$ | $\Sigma_-(\Sigma_+)^a$ | $\varepsilon_+(\Sigma_+)^b$ | $\varepsilon_-(\Sigma_+)^c$ | *Kasner* |

$$^a\Sigma_- = \tfrac{1}{\sqrt{3}}\sqrt{1 - 9\Sigma_+{}^2}\,.$$
$$^b\varepsilon_+ = \tfrac{1}{3}\Sigma_+(6\Sigma_+ + 1) - \tfrac{1}{9}\,.$$
$$^c\varepsilon_- = -\tfrac{\sqrt{3}}{9}(6\Sigma_+ - 1)\sqrt{1 - 9\Sigma_+{}^2}\,.$$

To get a clue of what goes on locally in an inhomogeneous universe with an effective cosmological constant we plot in Fig. 1 five significant cases of several numerical integrations of system (1). In Fig. 1 (left-up) we plot $\Theta/\sqrt{3\Lambda}$, the upper bound (3) locally representing the open de Sitter universe, the $\Theta = \sqrt{3\Lambda}$ line representing the flat de Sitter, and the bound (4) representing a closed de Sitter. In the strip $|\Theta| \leq \sqrt{3\Lambda}$ $^{(3)}R > 0$. The dynamically normalized, dimensionless 3-curvature scalar $^{(3)}R/\Theta^2$ is plotted in the right-upper corner. In the lower part we plot $\Omega = \Omega_M + \Omega_V$ and the deceleration parameter $q = -3\dot{\Theta}/\Theta^2 - 1$: one has $q < 0$ during inflation, $q < -1$ during super-inflation, and $q \to 2$ during collapse. When the de Sitter solution is asymptotically approached $\Omega \to 1$ and $q \to -1$. We see that by and large for inflationary evolutions $\Theta/\sqrt{3\Lambda}$ gets below the flat de Sitter line, then approaching it from below, asymptotically going as (4), with vanishing positive 3-curvature, $^{(3)}R \to 0^+$. Related to this, we have observed in our numerical results that $^{(3)}R$ does not change sign during the evolution, although we cannot formulate this as a rule; also, note that although $^{(3)}R \to \infty$ in the collapse, $^{(3)}R/\Theta^2 \to 0$: the approach to the singularity is Weyl dominated. Also, we observe that in general $\Omega$ does not approach unity even during an inflationary phase, unless super-inflation occurs.

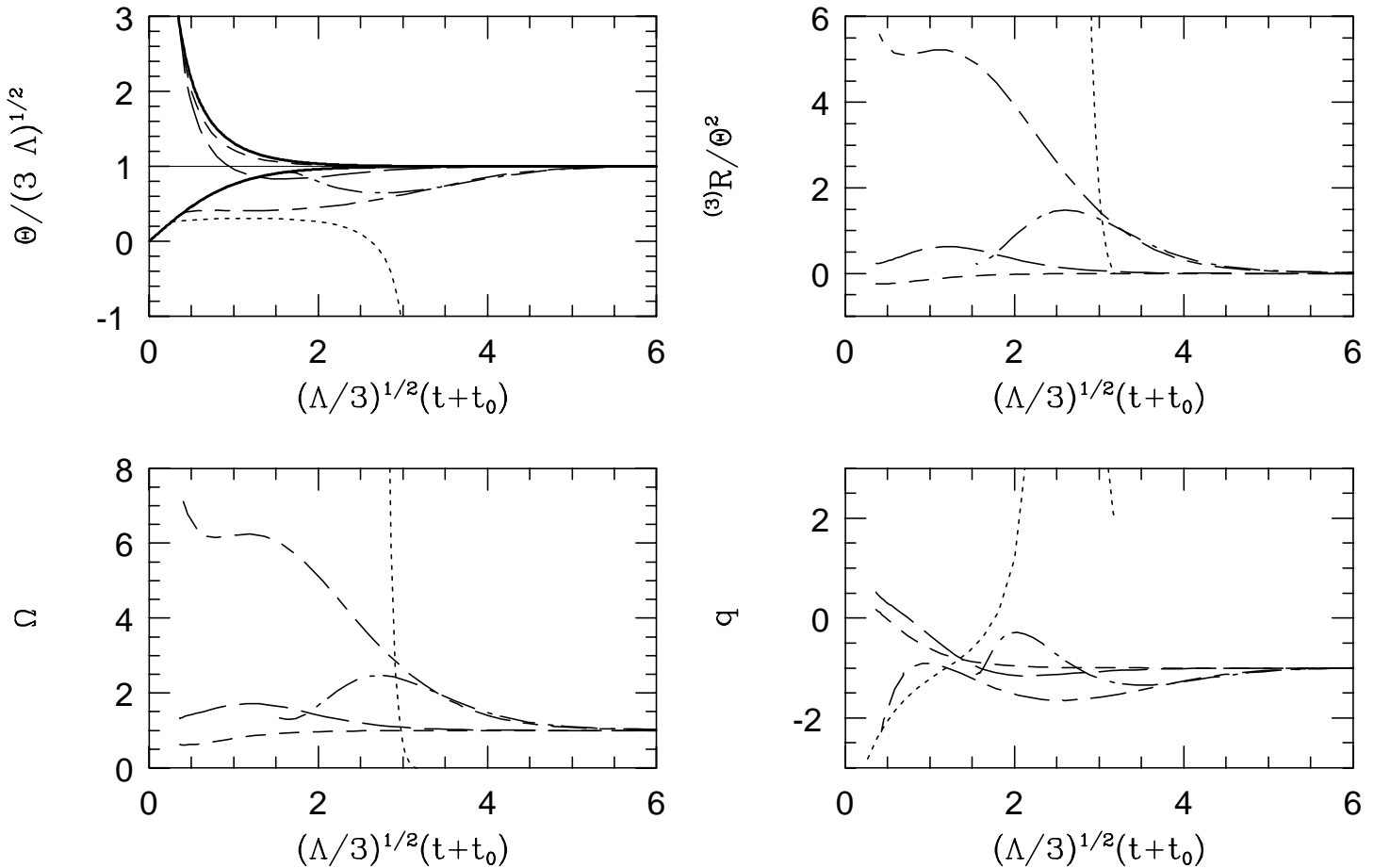

FIG. 1. Evolution of five inflationary cases, with one finally re-collapsing, against $\sqrt{\Lambda/3}(t+t_0)$, where $t_0$ and the upper bounds (thick lines) can be read from (3) and (4). For the re-collapsing case, before to diverge at turn around $(\Theta = 0)$, $^{(3)}R/\Theta^2$ and $\Omega$ are out of scale.

In this Letter we analyzed $H_{ab} = \omega_{ab} = p = 0$ models with a cosmological constant $\Lambda > 0$ thought as representing the local value of the vacuum energy density, and possibly causing the occurrence of an inflationary phase in certain patches of the universe with suitable initial conditions. Of course in a realistic scenario such a cosmological constant will ultimately decay into radiation, reheating the universe to a standard FRW phase. The picture that emerges is that of a generally inhomogeneous universe, with large patches of it where, thanks to an early inflationary evolution, the local properties are close to that of a flat FRW model. In this picture, however, the observable parameters $H_0$, $\Omega_0$ and $q_0$ should be thought as local values, not to be interpreted as giving the global properties of the universe.

MB would like to thank British PPARC (grant GR/J 36440) and Università degli Studi di Trieste for financial support, and SISSA for hospitality during the preparation of this work. SM and OP acknowledge Italian MURST for financial support.

# REFERENCES

[1] A. D. Linde, *Inflation and Quantum Cosmology* (Academic Press, new York, 1990); E. W. Kolb and M. S. Turner, *The Early Universe*, (Addison - Wesley, 1990).

[2] M. S. Turner, preprint FERMILAB-Conf-92/313-A.

[3] S. W. Hawking and I. G. Moss, Phys. Lett. B **110**, 35 (1982); W. Boucher and G. W. Gibbons, in *The Very Early Universe*, Eds. G. W. Gibbons and S. W. Hawking (Cambridge University Press, Cambridge, 1983); J. D. Barrow, *ibid*.

[4] R. M. Wald, Phys. Rev. D **28**, 2118 (1983).

[5] A. A. Starobinski, JETP Lett. **37**, 66 (1983); L. G. Jensen and J. A. Stein-Schabes, Phys. Rev. D **35**, 1146 (1987).

[6] D. S. Goldwirth and T. Piran, Phys. Repts. **214**, 223 (1992).

[7] We use units $c = 8\pi G = 1$, and definitions as in [17,11].

[8] A. Barnes and R. R. Rowlingson, Class. Quantum Grav. **6**, 949 (1989).

[9] S. Matarrese, O. Pantano and D. Saez, Phys. Rev. D **47**, 1311 (1993).

[10] S. Matarrese, O. Pantano and D. Saez, Phys. Rev. Lett. **72**, 320 (1994); Mon. Not. R. Astr. Soc. (to be published).

[11] M. Bruni, S. Matarrese and O. Pantano, SISSA preprint 85/94/A, (astro-ph/9406068).

[12] G. F. R. Ellis, in *Gravitation*, Eds. R. Mann and P. Wesson, (World Scientific 1991); P. Coles and G. F. R. Ellis, Nature (to be published).

[13] P. Szekeres, Comm. Math. Phys. **41**, 55 (1975); J. D. Barrow and J. Stein-Schabes, Phys. Lett. A **103**, 315 (1984).

[14] Physically equivalent replicas of these degenerate models are obtained setting $\sigma_- = \pm 3\sigma_+$ and $E_- = \pm 3E_+$.

[15] The expansion scalar $\Theta$, the shear $\sigma_{ab}$ and the electric Weyl tidal field $E_{ab}$ are defined with respect to this 4-velocity vector; see [17] and [11] for more details.

[16] Y. Kitada and K. Maeda, Class. Quantum Grav. **10**, 703 (1993).

[17] G. F. R. Ellis, *in General Relativity and Cosmology*, ed. R. K. Sachs (Academic Press, New York, 1971).

[18] F. Lucchin and S. Matarrese, Phys. Lett. B **164**, 282 (1985).

[19] We have used Maple V to find the stationary points, and to compute the eigenvalues of the Jacobian.

[20] Similar variables where defined by J. Wainwright and L. Hsu, Class. Quantum Grav. **6**, 1409 (1989).